\documentclass[eqsecnum,aps,preprint,epsf]{revtex4}
\usepackage{graphicx}
\usepackage{wrapfig}
\font\tenbifull=cmmib10 scaled 1200 
\font\tenbimed=cmmib9
\font\tenbismall=cmmib7
\def\bmit{\fam9 }
\mathchardef\bbkappa="7114
\mathchardef\bbrho="711A
\mathchardef\bbsigma="711B
\mathchardef\bbtau="711C
\mathchardef\bbvarrho="7125
\mathchardef\bbvarsigma="7126
\mathchardef\bbPhi="7008
\mathchardef\bbxi="7118

\def\boldrho{{\bmit\bbrho}}

\def\boldtau{{\bmit\bbtau}}

\def\boldPhi{{\bmit\bbPhi}}

\newcommand {\la} {\left\langle}
\newcommand {\ra} {\right\rangle}
\newcommand{\be}{\begin{eqnarray}&&}
\newcommand{\ee}{\end{eqnarray}}

\begin{document}
\title{Di-Electrons from Resonances in Nucleon-Nucleon Collisions}
\author{L.P. Kaptari}
\altaffiliation{On leave of absence from
Bogoliubov Lab. Theor. Phys. 141980, JINR,  Dubna, Russia}
\author{ B. K\"ampfer}
\affiliation{Forschungszentrum Dresden-Rossendorf, PF 510119, 01314 Dresden, Germany}

\begin{abstract}
The contribution of the low-lying nucleon resonances
$P_{33}(1232)$, $P_{11}(1440)$  $D_{13}(1520)$ and $S_{11}(1535)$
to the invariant mass spectra of di-electrons stemming from the
exclusive processes $pp\to pp \ e^+e^-$  and $pn\to pn \ e^+e^-$
is investigated  within a fully covariant and  gauge invariant diagrammatical
approach. We employ, within the one-boson exchange approximation,
effective nucleon-meson interactions including the exchange
mesons $\pi$, $\eta$,  $\sigma$, $\omega$ and $\rho$ as well as excitations and radiative
decays of the above low-lying nucleon resonances. 
The total contribution of these resonances
is dominant, however, bremsstrahlung processes in $pp$ and, 
in particular, $pn$ collisions at beam energies of 1 - 2 GeV 
are still significant in certain phase space regions.
\end{abstract}

\maketitle

\section{Introduction}

The experimental study of di-electrons as penetrating probes in relativistic
heavy-ion collisions is aimed at identifying medium modifications of hadrons,
in particular of the vector mesons $\rho$, $\omega$ and $\phi$ \cite{Rapp_Wambach}.
Previous measurements of di-electrons in the reaction ${}^{12}C + {}^{12}C$ 
at kinetic beam energy of 1.04 AGeV performed by
the DLS collaboration \cite{DLS} have been confirmed
recently by the HADES collaboration \cite{HADES1}, 
at least in phase space regions covered by both experiments. 
Various transport models have been employed \cite{brat08,also}
for understanding and interpreting the di-electron data \cite{DLS,HADES1,HADES2}.
Among the important sources for di-electrons in the low-mass region
are $\pi^0$, $\Delta$ and $\eta$ Dalitz decays and bremsstrahlung as well
\cite{brat08,also}.

The elementary cross section
for virtual nucleon-nucleon bremsstrahlung with $\gamma^* \to e^+ e^-$
as a subprocess in heavy-ion collisions was parameterized often within 
the soft-photon approximation (cf.\ \cite{tormoz}), 
which is appropriate at low kinetic energies, 
where the photon is quasi-real, but becomes questionably at higher energies and
at higher virtualities of the $\gamma^*$. 
Moreover, the soft-photon approximation preserves only 
approximately the gauge invariance, 
and the violation of gauge invariance increases with initial energy.  
In Ref.~\cite{ourbrem}, based on previous
investigations \cite{ourPhi,ourOmega,titov,mosel_calc,Kapusta}, a fully covariant and
gauge invariant approach has been proposed to parameterize 
the bremsstrahlung amplitude in elementary $pp$ and $pn$ collisions.
It was demonstrated  that, in order to preserve the gauge invariance in $pn$ reactions,
one has to include additional diagrams
with meson exchange currents and, for the couplings with field derivatives,
to introduce  contact terms, the so-called seagull or Kroll-Rudermann \cite{KR}  
type diagrams. The resulting $pn$ bremsstrahlung cross section was found to
essentially differ from the one obtained within previous quasi-classical calculations.
(This conclusion has been confirmed in Ref.~\cite{shyam08}.) 
The calculations reported in \cite{brat08} utilized
the bremsstrahlung cross sections of \cite{ourbrem} and, 
indeed, are capable describing perfectly
the  DLS \cite{DLS} and the recent HADES di-electron data \cite{HADES1,HADES2}
for the reaction ${}^{12}C + {}^{12}C$.
Hence, one can assert that the so-called ''DLS puzzle''
originated from scarce knowledge of elementary cross sections 
used in transport models, in particular the elementary
nucleon-nucleon bremsstrahlung. 

In covariant approaches, based on an effective meson-nucleon theory to calculate
the bremsstrahlung of di-electrons from nucleon-nucleon scattering,
the effective parameters have been adjusted to describe
elastic nucleon-nucleon ($NN$) and inelastic $NN \to NN \pi$
processes at intermediate energies. Excitations of resonances
have been studied at the same time, 
and it is found that at intermediate energies the main contribution
comes from $\Delta$ resonances
(see also Ref.~\cite{ernst}), whereas excitations of higher mass
resonances are often  neglected.
The role of higher mass and spin nucleon resonances
at energies near the vector meson ($\rho$, $\omega$ and $\phi$)
production thresholds have been investigated
for proton-proton collisions in several papers 
(see, e.g., Refs.~\cite{nakayama,fuchs} and further references therein)
with the conclusion that at threshold-near energies the inclusion
of heavier resonances also leads to a good description of data.
However, as demonstrated in Refs.~\cite{nakayama,ourOmega}
calculations with a reasonable readjustment of the effective parameters
can equally well describe the data without higher mass and spin resonances.
In contrast, for di-electron production in photon and pion
induced reactions, excitations of low-lying as well as
heavier resonances can play a role \cite{lutzNew}.

In the present paper we investigate in some detail the role of nucleon resonances 
with masses close to the $\Delta$ for di-electron production in $NN$ collisions. 
Besides the $\Delta$ we consider the low-lying
$P_{11}(1440)$, $D_{13}(1520)$ and $S_{11}(1535)$ resonances  which are expected
to contribute at larger values of the $e^+e^-$ invariant mass and, therefore, can
modify the shape of the $e^+e^-$ mass distribution at the kinematical limit.

Our paper is organized as follows. In section \ref{subsection1} we recall the kinematics and
the general expressions for the cross section. 
The purely electromagnetic part of the cross section
is considered in section \ref{sub0}, where the integration over the leptonic variable is
performed analytically and an expression for the cross section is presented.
In sections \ref{subs1} and \ref{subsecGauge}, the effective Lagrangians and the problem
of gauge invariance within the one-boson exchange model are  discussed.
Meson exchange diagrams and
seagull terms are considered in this section as well. Results for
the invariant-mass distribution of di-electrons 
stemming from $pp$ and $pn$ bremsstrahlung processes, where
only nucleons and mesons are involved, 
are reported in section \ref{subsecTri}.
The role of resonances is investigated in section \ref{reson}.
In section \ref{isobar}, the contribution of the $\Delta$ isobar is considered.
In particular, the choice of the coupling constants together with the off-mass 
shell parameters is discussed. 
A comparison of the contributions from bremsstrahlung and $\Delta$ is presented
also in this subsection. The nucleon resonances with spin 1/2 and 3/2 are considered
in sections \ref{spin05} and \ref{spin15}, respectively. 
The adjustment of effective parameters to experimental
data and the parametrization of the energy dependence 
of the resonance widths are reported in detail.
The individual contributions of each resonance are analyzed. 
The total cross section as a coherent sum of bremsstrahlung 
and resonance contributions, including all interference effects, 
is presented for two experimentally relevant
kinetic energies in $pp$ and $pn$ collision. 
The summary and conclusions can be found in section \ref{summary}.

\section{Di-electrons from \mbox{\boldmath $NN $} collisions}
\label{section1}

\subsection{Kinematics and Notation}  \label{subsection1}

We consider the exclusive $e^+e^-$ production in $NN$ reactions of the type
\be
N_1 (P_1) + N_2 (P_2)  \to N_1' (P_1') + N_2' (P_2') + e^+(k_1) + e^-(k_2).
\label{reac1}
\ee
The invariant eight-fold cross section is
\begin{eqnarray} 
d^8\sigma &=&
\frac{1}{2\sqrt{\lambda(s,m^2,m^2)}}\frac{1}{(2\pi)^8}
\frac14\sum\limits_{spins} \,
|T |^2 \,  \frac{1}{n! } d s_{12} ds_\gamma \\
&\times&  dR_2(P_1+P_2\to q +P_{12})\
dR_2(q\to k_1+k_2)\ dR_2(P_{12}\to P_1'+P_2')\ ,
\nonumber
\label{sechenie}
\end{eqnarray}
where the two-body invariant phase space volume $R_2$ is defined as
\begin{equation}
dR_2(a+b \to c+d) = d^4 P_c\ d^4P_d\ \delta^{(4)}(P_a+P_b-P_c-P_d)\
\delta (P_c^2-m_c^2)\ \delta(P_d^2-m_d^2).
\label{spase}
\end{equation}
The four-momenta of initial  ($P_1, P_2$) and  final  ($P_1', P_2'$) nucleons are
$P=(E_{\bf P},{\bf P})$ with $E_{\bf P}=\sqrt{m^2+{\bf P}^2}$; an  analogous notation
is used for the lepton momenta $k_{1,2}$;
$m$ denotes the nucleon mass, while the electron mass
can be neglected for the present kinematics.
The invariant mass of two particles is  denoted hereafter as $s$ with $s=(P_1+P_2)^2$;
along with this notation  for the invariant mass
of the virtual photon throughout the paper we also use
the more familiar notation $q^2$  with $q^2 \equiv M^2$.
The kinematical  factor $\lambda$ is
$\lambda(x^2,y^2,z^2)=(x^2-(y+z)^2)(x^2-(y-z)^2)$;
the factor $1/n!$ accounts for $n$ identical
particles in the final state.

\subsection{Leptonic tensor}\label{sub0}

The di-electron production process is considered as decay of a virtual photon
produced in strong and electromagnetic $NN$ interactions
from different elementary reactions, e.g.,
bremsstrahlung, Dalitz decay, vector meson decay etc.\ \cite{brat}.
For such a process the general expression for the
invariant amplitude squared reads
\be
|T|^2=W_{\mu\nu} \, \frac{e^4}{q^4}  l^{\mu\nu},
\label{ophe}
\ee
where the momentum  of the virtual photon is denoted as
$q  =(k_1+k_2)$;  $e$ is the elementary electric charge.
The purely electromagnetic decay vertex of the virtual photon is
determined by the leptonic tensor
$ l^{\mu\nu} =\sum\limits_{spins} j^\mu j^\nu$
with the current
$j^\mu = \bar u(k_1,s_1)\ \gamma^\mu v (k_2,s_2)$,
where $\bar u$ and $v$ are the corresponding Dirac bispinors for the outgoing
electron and positron. 
The leptonic tensor reads explicitly
\be
l_{\mu\nu} = 4 \left( k_{1\mu} k_{2\nu}+  k_{1\nu} k_{2\mu}
-g_{\mu\nu} (k_1\cdot  k_2)\right)
\ee
for unpolarized di-electrons. 

The integral over the leptonic phase space is easily calculated due to
its covariance and the fact that the only "external" variable on which it 
can depend is the di-electron four-momentum $q$,
\begin{equation}
\int l_{\mu\nu}(k_1,k_2,q) dR_2(q\to k_1+k_2)=\frac{2\pi}{3}q^2
\left(-g_{\mu\nu}+\frac{q_\mu q_\nu}{q^2}\right).
\label{intehral}
\end{equation}
Obviously, in virtue of gauge invariance of the
electromagnetic tensors, $q_\mu l^{\mu\nu} = q_\nu l^{\mu\nu}
= q^\nu W_{\mu\nu} = q^\nu W_{\mu\nu}=0$, only the first term in the 
r.h.s.\ of Eq.~(\ref{intehral})
contributes, so that we obtain
\begin{eqnarray}
\frac{d\sigma}{d M}=
-\frac{\alpha_{em}^2}{6 Ms (4\pi)^5}
\int d s_{12} d\Omega_\gamma^* d\Omega_{12}^*
\sqrt{\displaystyle\frac{\lambda(s,s_{12},M^2)\lambda(s_{12},m^2,m^2)}
{ s_{12}^2\lambda(s,m^2,m^2)}} \sum\limits_{spins} J_\mu J^{  +\mu},
\label{diffcross}
\end{eqnarray}
where $d\Omega_\gamma^*$
and $d\Omega_{12}^*$ are defined in the center of mass of initial 
and final nucleons, respectively;
$\alpha_{em}$ stands for the electromagnetic fine structure constant.

\subsection{Lagrangians and parameters} \label{subs1}

The covariant hadronic current $J_\mu$ is evaluated
within a meson-nucleon theory based on effective interaction Lagrangians
which consist on two parts describing the strong and electromagnetic
interaction. In our approach, the strong interaction
among nucleons is mediated by four exchange mesons:
scalar ($\sigma$), pseudoscalar-isovector  ($\pi$),
and neutral vector ($\omega$) and vector-isovector  ($\rho$)  mesons
\cite{mosel_calc,ourOmega,ourPhi,bonncd}.
We adopt the nucleon-nucleon-meson (NNM)  interaction terms
\begin{eqnarray}
{\cal L}_{NN\sigma}&=& g_\sigma \bar N  N \it\Phi_{(\sigma)} , \\
{\cal L}_{ NN\pi}&=&
-\frac{f_{ NN\pi}}{m_\pi}\bar N\gamma_5\gamma^\mu  \boldtau 
(\partial_\mu {\boldPhi_{(\pi)}}) N , \label{pseudo_vector}\\
{\cal L}_{ NN\rho}&=&
-g_{ NN \rho}\left(\bar N \gamma_\mu{\boldtau} N
\boldPhi_{(\rho)}^\mu-\frac{\kappa_\rho}{2m}
\bar N\sigma_{\mu\nu}{\boldtau}N\partial^\nu \boldPhi_{(\rho)}^\mu \right) , \label{rhonn}\\
{\cal L}_{ NN\omega}&=&
-g_{  NN \omega}\left(
\bar N \gamma_\mu N {\it\Phi}_{(\omega)}^\mu-
\frac{\kappa_{\omega}}{2m}
\bar N \sigma_{\mu\nu}  N \partial^\nu \it\Phi_{(\omega)}^\mu\right),
\label{mnn}
\end{eqnarray}
where $N$ and $\it\Phi_{(M)}$ denote the nucleon and meson fields, respectively,
and bold face letters stand for isovectors.
All couplings with off-mass shell
particles are dressed by monopole form factors
$F_M=\left(\Lambda^2_M-\mu_M^2\right)/\left(\Lambda^2_M-k^2_M\right)$,
where $k^2_M$ is the four-momentum of a virtual meson
with mass $\mu_M$.
The effective parameters and their dependence   
on the initial energy are adjusted to experimental data
on $N N$ scattering at the considered intermediate energies \cite{mosel_calc,ikh_model}.

\subsection{Gauge invariance}\label{subsecGauge}

\begin{figure}[h]  
\vskip -6mm
\hspace*{-0.5cm}\includegraphics[width=0.95 \textwidth]{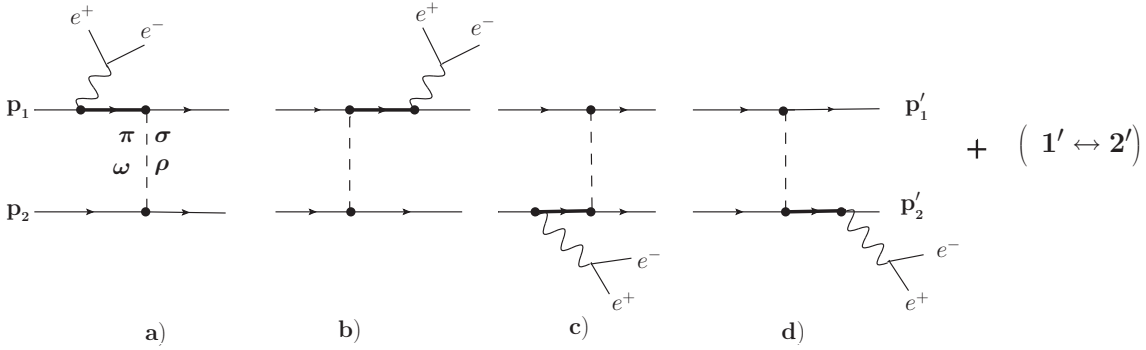} %
\caption{ Bremsstrahlung diagrams for the process $N_1+N_2\to N_1'+N_2'+e^+ +e^-$.
Fat lines are for resonances considered in section \ref{reson}.}
\label{fig1}
\end{figure}

\begin{figure}[h]  
\vskip -9mm
\includegraphics[width=1.\textwidth]{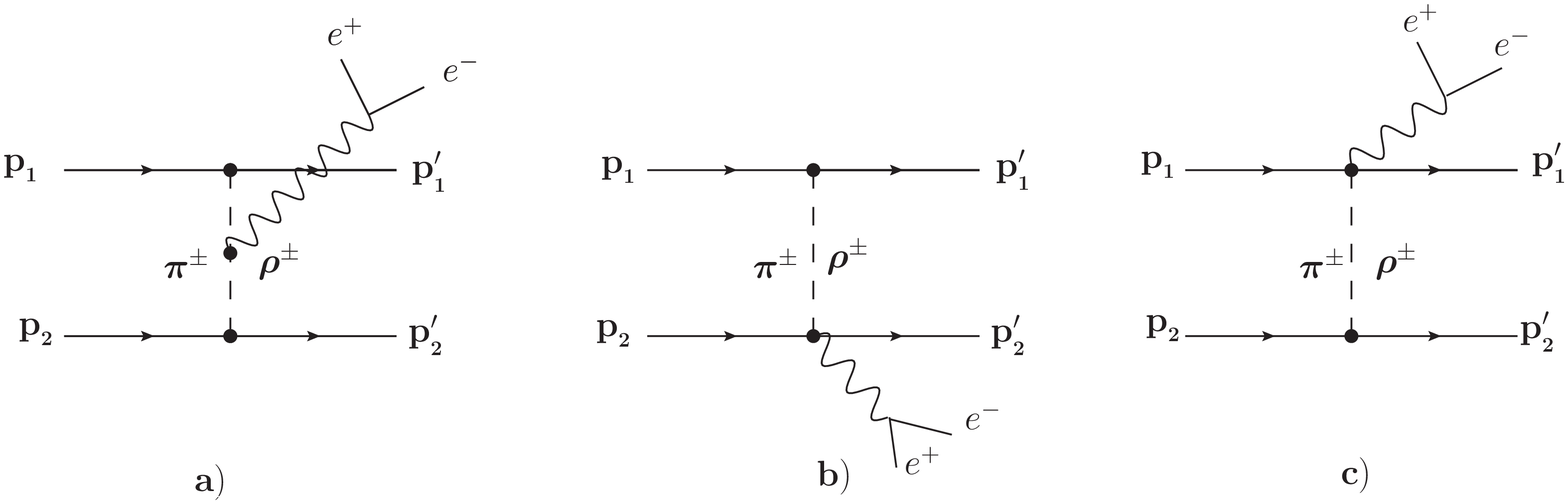} %
\caption{ Contribution of meson exchange currents (a)
and seagull terms (b, c)
to the  process  $N_1+N_2 \to N_1' + N_2' + e^+ e^-$,
where $N_1$ and  $N_2'$ stand for protons and
$N_2$ and $N_1'$ denote neutrons.}
\label{fig2}
\end{figure}

The form of the cross section Eq.~(\ref{diffcross}) exploits
essentially  the gauge invariance of hadronic and leptonic
tensors. This implies that in elaborating  models
for the reaction (\ref{reac1}) with effective Lagrangians,
particular attention must be devoted to
the gauge invariance of the computed currents with the mandatory condition $q_\mu J^\mu=0$.
In our approach, i.e., in
the one-boson exchange approximation (OBE) for the strong $NN$ interaction
and one-photon exchange for the electromagnetic
production of $e^+e^-$, the current $J_\mu$ is determined by diagrams of two types:
(i) the ones which describe the creation of a virtual photon
with $q^2>0$ as pure nucleon bremsstrahlung as depicted in Fig.~\ref{fig1} and
(ii) in case of exchange of charged mesons, the emission of a virtual  photon ($\gamma^*$)
from internal meson lines, see Fig.~\ref{fig2}a.
For these diagrams the gauge invariance is tightly connected with the
two-body Ward-Takahashi (WT) identity
\be
q_\mu \Gamma^\mu (p',p) = \frac{e(1+\tau_3)}{2}\left(
S^{-1}(p') - S^{-1}(p)\right),
\label{wt}
\ee
where $\Gamma^\mu$ denotes the electromagnetic vertex and
$S(p)$ is the (full) propagator of the respective particle. It is straightforward to
show that, if (\ref{wt}) is to be fulfilled, then
pairwise two diagrams with exchange of neutral
mesons and  pre-emission and post-emission of $\gamma^*$ (cf.\  Figs.~\ref{fig1}a) and b))
cancel each other, hence ensuring $q^\mu J_\mu=0$, i.e.,
current conservation.
This is also true  after dressing the vertices with phenomenological
form factors. However, in case of charged meson exchange
the WT identity is not any more automatically fulfilled. This is
because the nucleon momenta are interchanged
and, consequently,  the "right" and "left" internal nucleon propagators are defined
for different momenta of the exchanged meson.

In order to restore the gauge invariance on this level
one must consider additional diagrams with emission of the
virtual photon by the charged meson exchange (Fig.~\ref{fig2}a)
which exactly compensates
the non-zero part of the current divergence, and thus gauge invariance is restored.
This holds true for bar $NNM$ vertices without cut-off form factors.
Inclusion of additional form factors again leads to non-conserved
currents. There are several prescriptions of how
to preserve gauge invariance within effective theories with
cut-off form factors.

\begin{figure}[h]  %
\includegraphics[width=.9\textwidth]{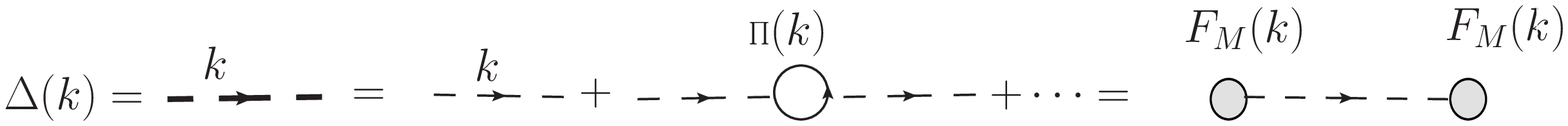} %
\caption{Graphical illustration  of the cut-off form factor as
self-energy corrections to the full propagator \cite{gross}.}
\label{gross}
\end{figure}

The main idea of these
prescriptions is to consider the cut-off form factors as phenomenological part
of the self-energy corrections to the corresponding propagators;
the full propagators are to be treated
as the bare ones multiplied at  both ends by a form factor \cite{gross}.
Then, the full propagator, e.g. for mesons, can be defined as
\be
\Delta(k)=\frac{1}{k^2-\mu_M^2+\Pi (k)}\equiv\frac{F_M^2(k)}{k^2-\mu_M^2},
\ee
where $\Pi (k)$ is the self-energy correction (see Fig.~\ref{gross}).

In the simplest case, for  mesonic vertices with pseudoscalar couplings,
the bare mesonic vertex $\Gamma_\mu^M=\left( k_{1\mu}+k_{2\mu}\right)$
receives an additional factor \cite{schafer,mathiot,ourbrem} becoming
\be
\Gamma_\mu^{\gamma M}=\left( k_{1\mu}+k_{2\mu}\right)
\frac{\left(\Lambda_M^2 - k_1^2\right)}{\left(\Lambda_M^2 - \mu_M^2\right)}
\frac{\left(\Lambda_M^2 - k_2^2\right)}{\left(\Lambda_M^2 - \mu_M^2\right)}
\left( 1-\frac{k_1^2-\mu_M^2}{\Lambda_m^2-k_2^2}
-\frac{k_2^2-\mu_M^2}{\Lambda_M^2-k_1^2}\right ).
\label{pionem}
\ee

The above prescriptions for restoration of the gauge invariance
in $pn$ collisions are valid only for pion exchange diagrams with the
interaction vertices independent of the momentum $k$ of the exchanged meson, i.e.,
solely for the case of pseudo-scalar $\pi NN $ coupling.
The presence of field derivatives in the interaction Lagrangian, e.g. the case
for pseudo-vector  $\pi NN $ coupling or for vector mesons, 
Eqs.~(\ref{pseudo_vector}) and (\ref{rhonn}),
requires a more refined treatment of the gauge invariance.
In the simplest case, besides the WT identity condition (with full propagators)
for the diagram \ref{fig2}a,
the gauge invariance requires an introduction of covariant derivatives, 
i.e. the replacement of the partial derivatives,
including the $NN M$ vertices, by a covariant form (minimal coupling).
Such a procedure generates another kind of Feynman diagrams with
contact terms, i.e., vertices with four lines, known also as
Kroll-Rudermann \cite{KR} or seagull like diagrams, see Figs.~\ref{fig2}b and c.
We include therefore in our calculations these diagrams by the corresponding
interaction Lagrangian
\begin{equation}
{\cal L}_{ NN\pi\gamma} =
-\frac{\hat e f_{ NN\pi}}{m_\pi}\bar N\gamma_5\gamma^\mu A_\mu
({\boldtau \boldPhi_{(\pi)}})N
\end{equation}
with electromagnetic four-potential $A_\mu$ and charge operator $\hat e$ of the pion.
Analogously for the $\rho NN$ coupling one has to replace
\begin{equation}
\frac{\kappa_\rho}{2m}
\bar N\sigma_{\mu\nu}{\boldtau} N \partial^\nu \boldPhi_{(\rho)}^\mu\,
\longrightarrow \,  \frac{\kappa_\rho}{2m}
\bar N\sigma_{\mu\nu}{\boldtau}N\partial^\nu \boldPhi_{(\rho)}^\mu
+\hat e \frac{\kappa_\rho}{2m}
\bar N\sigma_{\mu\nu}{\boldtau}N A_\nu \boldPhi_{(\rho)}^\mu.
\label{sqerho}
\end{equation}
Gauge invariance is henceforth ensured.
It should be stressed that as far as the tensor
part of the $\rho NN$ Lagrangian (see Eq.~(\ref{rhonn})) is accounted for, 
the prescription (\ref{sqerho})
must be  mandatorily applied, regardless of the choice of $\pi NN$ coupling. 
This implies that calculations with pseudo-scalar couplings for the pion-nucleon vertex 
violates gauge invariance due to $\rho$ meson exchange. 
Our numerical calculations show that the effect
of such $\rho$ exchange seagull type diagrams vary from $10\%$ at 
low di-electron invariant masses up to 35\% at the kinematical limit.

All electromagnetic $NN\gamma$ vertices correspond to the interaction
Lagrangian
\begin{eqnarray}
{\cal L}^{em}_{NN\gamma}= -e\left( \bar N\gamma_\mu N \right) A^\mu+
e\kappa \bar N \left( \frac{\sigma_{\mu\nu}}{4m}{\cal F}^{\mu\nu}\right) N
\label{elm}
\end{eqnarray}
with the field strength tensor
${\cal F}_{\mu\nu}=\partial_\nu A_\mu - \partial_\mu A_\nu$,
and $\kappa$ as the anomalous magnetic moment of the nucleon
($\kappa=1.793$ for  protons and $\kappa=-1.913$ for neutrons).

\subsection{Results for bremsstrahlung} \label{subsecTri}

\begin{figure}[h]  %
\vskip -9mm
\includegraphics[width=1.\textwidth]{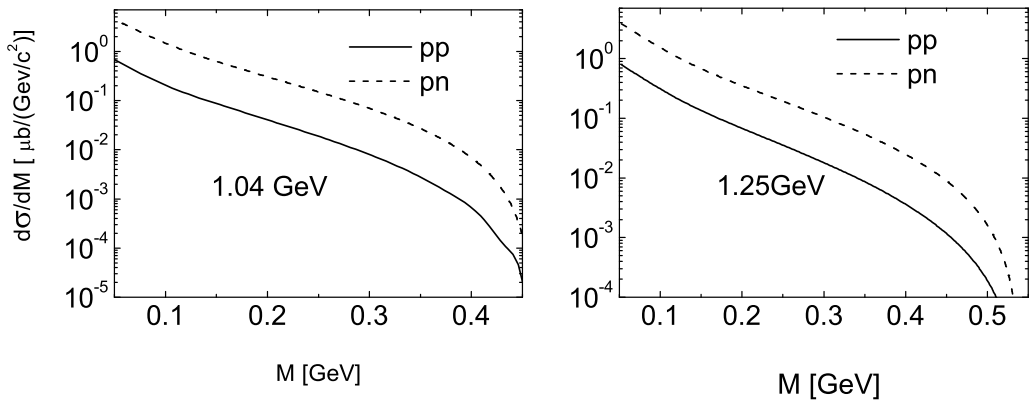} %
\vskip -9mm
\caption{Contribution of the bremsstrahlung diagrams in
Figs.~\ref{fig1} and \ref{fig2} (without nucleon resonances)
to the $e^+e^-$
invariant mass distribution at two kinetic energies 
(left: 1.04~GeV, right: 1.25~GeV).
Solid lines correspond to $pp$ reactions; 
dashed lines are for $pn$ reactions.}
\label{fig3}
\end{figure}

The OBE parameters and their energy dependence
have been taken as in Ref.~\cite{ourbrem,mosel_calc}. Figure \ref{fig3}
exhibits results of our calculations of the invariant-mass distribution of di-electrons
in $pp$ and $pn$ collisions from bremsstrahlung
processes in Figs.~\ref{fig1} and \ref{fig2} (nucleons only)
at two values of the kinetic energy,
1.04~GeV and 1.25~GeV as relevant for DLS \cite{DLS} and HADES \cite{HADES1} measurements.
In our actual calculations
we include, besides the mentioned four exchange mesons 
$\pi$, $\sigma$, $\rho$ and $\omega$
also a ''counter term'' simulating a heavy axial vector-isovector meson,
with the goal to cancel singularities
of the pion potential at the origin \cite{mosel_calc}.
The dotted lines in Fig.~\ref{fig3}
depict the cross section in $pn$ collision, while the solid lines stand for
results of $pp$ reactions. It is seen that the $pn$ cross section is 
by a factor $5 - 6$ larger than the $pp$ cross section. 
(The situation for real photon emission is similar: 
The $pn$ channel has a significant contribution,
while, due to a destructive  interference, the $pp$ channel is much weaker and
is often neglected \cite{barz}.)
This is due to isospin effects for the charged $\pi$ and $\rho$ mesons 
and different interference effects in $pp$ and $pn$ channels. 
In $pn$ reactions, 
additional contributions stem from the emission off charged 
exchange mesons and from corresponding seagull type diagrams. 
Note that in both channels,
$pp$ and $pn$, numerical tests of gauge invariance
can serve as additional check of the code. The meson exchange diagrams together
with contact terms amplify the contribution of pure nucleonic currents;
their contribution  is of the order of $\sim 40\%$ at
low values of the invariant di-electron masses and increases 
up to a factor $2-3$ at higher invariant masses.

\section{resonances}\label{reson}

Intermediate baryon resonances play an important role
in di-electron production in $NN$ collisions at beam energies in the 1 - 2 GeV region
\cite{fuchs,ernst,nakayama,lutzNew,kapusta,mosel_calc,ikh_model,schafer,brat}.
The main contribution
to the cross section stems from the $\Delta$ isobar \cite{mosel_calc}.
Also, the low-lying nucleon resonances such as $N^*(1440) $, $N^*(1520)$ and
$N^*({1535})$   contribute to the cross section. We are going 
to investigate separately each of these resonances
represented in Fig.~\ref{fig1} by fat lines.

\subsection{\mbox{\boldmath $P_{33}(1232)$}}\label{isobar}

Since the isospin of the $\Delta$ is $3/2$
only the isovector mesons $\pi$ and $\rho$ couple to nucleons and $\Delta$'s.
The form of the effective $\Delta N$ interaction was thoroughly
investigated in literature in connection with $NN$
scattering \cite{bonncd,holinde,weise},
pion photo- and electroproduction \cite{pascQuant,photo,davidson,elctro,lenske}.
The effective Lagrangians of the $\Delta N M$ interactions
read \cite{holinde,weise,pascQuant}
\be
{\cal L}_{\Delta N\pi} = \frac{f_{\Delta N\pi}}{m_\pi}
\left(\bar\Psi_\Delta^\alpha \ {\bf T} \ \partial_\alpha \boldPhi_{(\pi)} N
\right) +h.c., \label{pionDelta}\\
&& {\cal L}_{\Delta N\rho} =\frac{if_{\Delta N \rho}}{m_\rho}
\left(\bar\Psi_{\Delta\,\alpha} {\bf T} 
\left\{ \partial^\beta\boldPhi_{(\rho)}^\alpha-
\partial^\alpha\boldPhi_{(\rho)}^\beta\right\}\gamma_\beta\gamma_5 N\right)
+h.c.
\label{rhoDelta}
\ee
with $f_{\Delta N\pi}=2.13 $ GeV and $f_{\Delta N\rho}=7.14$~GeV \cite{schafer}.
The $\Delta N M$ vertices are dressed by cut-off form factors
\be
F^{\Delta  N M} = \left[ \frac{\Lambda_{\Delta  N M}^2-\mu_M^2 }
{\Lambda_{\Delta  N M}^2-k^2}\right]^2,
\label{cutDelta}
\ee
where $\Lambda_{ \Delta N \pi}=1.4214$~GeV and
$ \Lambda_{ \Delta N\rho}=2.273$~GeV \cite{schafer}.
The symbol ${\bf T}$ in Eqs.~(\ref{pionDelta}) and (\ref{rhoDelta})
stands for the isospin transition matrix, and
$\Psi_\Delta$ denotes  the field describing the
$\Delta$.  Particles with  higher spins ($s>1$)
are treated usually within the Rarita-Schwinger formalism in accordance with which
the $\Delta$ propagator has the form
\be
S_\Delta^{\alpha\beta}(p,m_\Delta) =
\frac{i\left(\hat p + m_\Delta\right)}{p^2 - m_\Delta^2}  \
P_{\frac32}^{\alpha\beta}(p,m_\Delta)
\label{raritaprop}
\ee
 with the spin projection operator $P_{\frac32}^{\alpha\beta}(p)$ defined as
\be
P_{\frac32}^{\alpha\beta}(p,m_\Delta) =-g^{\alpha\beta} +\frac13 \gamma^\alpha\gamma^\beta
+\frac{2}{3m_\Delta^2} p^{\alpha}p^\beta + \frac{1}{3m_\Delta}
\left( \gamma^\alpha p^\beta - \gamma^\beta p^\alpha\right).
\label{spinproj}
\ee
In addition, to take into account the widths
of $\Delta$, the mass in the denominator
of the scalar part of the propagator is modified as
$m_\Delta \to m_\Delta - i\Gamma_\Delta/2$.
For the kinematics considered here, the "mass" $\sqrt{p^2}$  of the
intermediate $\Delta$ can be rather far from its pole value. The
width, as a function of $p^2$, is calculated
as a sum of partial widths being dominated by  the one-pion ($\Delta\to N \pi$)
and two-pion ($\Delta\to N \rho\to N \pi\pi$) decay
channels \cite{shyamwidts}.

The general form of the $\Delta N \gamma $ coupling
satisfying  gauge invariance can be written
as \cite{davidson,feuster,pascTjon,pascScholten}
\begin{eqnarray}
{\cal L}_{\Delta N \gamma}&=& -i\frac{eg_1}{2m}
\bar\Psi^\alpha \Theta_{\alpha\mu}(z_1)\gamma_\nu
\gamma_5 {\bf T}_3
N {\cal F}^{\nu\mu}-  
\frac{eg_2}{4m^2}\bar\Psi^\alpha \Theta_{\alpha\mu}(z_2)\gamma_5
{\bf T}_3 \left( \partial_\nu N \right) {\cal F}^{\nu\mu} \nonumber\\
&& -\frac{eg_3}{4m^2}\bar\Psi^\alpha \Theta_{\alpha\mu}(z_3)\gamma_5
{\bf T}_3  N  \partial_\nu {\cal F}^{\nu\mu} +h.c., \label{deltaLag}\\
\Theta_{\alpha\mu}(z) &=& g_{\alpha\mu}+
[z +\frac12( 1+4z)A]\gamma_\alpha\gamma_\mu,
\label{theta}
\end{eqnarray}
where $A$ is a constant reflecting the invariance of the free $\Delta$
Lagrangian with respect to point transformations which, according to common
practice, is taken $A=-1$. The other parameters, $z_1,z_2$ and $z_3$ are  also connected
with point transformations; however, they characterize  the off-mass shell
$\Delta$ resonance and remain unconstrained.
The two coupling constants $g_1$ and $g_2$ in Eq. (\ref{deltaLag})
can be estimated from experimental data with real photons ($q^2=0$)
by evaluating the helicity amplitudes~\cite{helicity} of the process $\Delta\to\gamma N$
obtained within the Lagrangian Eq.~(\ref{deltaLag}). One finds explicitly
\begin{eqnarray} &&
eg_1\la {\bf T}_3  \ra=-2\sqrt{2} \frac{m}{m+m_\Delta}\sqrt{\frac{m_\Delta m}{|{\bf q}^*|}}
\left[ \sqrt{3}A_{\frac12}^{PDG}+A_{\frac32}^{PDG}\right],\label{g1}\\&&
eg_2\la {\bf T}_3  \ra=-4\sqrt{2} \left( \frac{m}{m_\Delta |{\bf q}^*|}\right)^{\frac32}
mm_\Delta
\left[ \sqrt{3}A_{\frac12}^{PDG}-\frac{m}{m_\Delta}A_{\frac32}^{PDG}\right],\label{g2}
\end{eqnarray}
where $|{\bf q}^*|$ is the three-momentum of the photon in the $\Delta$ center-of-mass system,
and the helicity amplitudes $A_{\frac12(\frac32)}^{PDG}$ are normalized in such a way
that they correspond to values listed by Particle Data Group (PDG) \cite{PDG}. 
It should be pointed out that
both coupling constants $g_{1,2}$ are quite sensitive to the values of
the helicity amplitudes, which, according to
PDG, vary in rather large intervals: 
$- A_{\frac12}^{PDG}\sim 0.128 \cdots 0.145$,
$- A_{\frac32}^{PDG}\sim 0.243 \cdots 0.261$, \cite{PDG} providing the uncertainties
$g_1 \sim 4.5 \cdots 5.5$ and $g_2 \sim 4.5 \cdots 8.5$.
In principle, the computed electromagnetic
width of $\Delta$ can serve as an additional constraint for the coupling constants.
However it turns out that it is less sensitive to the actual choice of
$g_{1,2}$ yielding satisfactorily good results for different sets of $g_{1,2}$
obtained from helicity amplitudes. 
Another uncertainty ($\sim 20\%$) in Eqs.~(\ref{g1}) and (\ref{g2})
follows from the normalization of the isospin transition matrix $\la {\bf T}_3\ra$
which can be chosen  either $\sqrt{\frac23}$ or $1$ 
(see discussion appendix A in \cite{ourbrem}).
The remaining constants in Eq.~(\ref{deltaLag}), $z_{1, 2 ,3}$ and $g_3$,
do not contribute in processes with on-mass shell particles and cannot be directly
related to data. As a matter of fact,  
they are considered as free fitting parameters \cite{feuster}.
In the present calculation we use $g_1=5.478$, $z_1=0.05$, $g_2=7.611$, $z_2=1.499$,
$g_3=7.0$ and $z_3=0$ being consistent with the available experimental data on helicity
amplitudes, electromagnetic decay width and pion photoproduction (cf.\ Ref.~\cite{feuster}).

\begin{figure}[h]  %
\vskip -9mm
\includegraphics[width=0.9\textwidth]{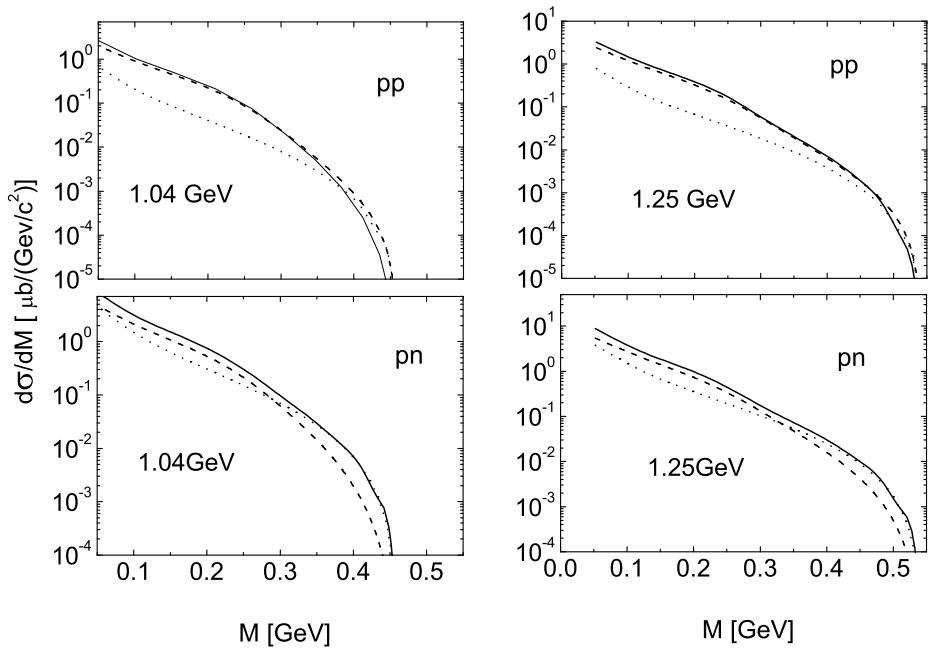} %
\vskip -9mm
\caption{Invariant mass distribution of $e^+e^-$ pairs 
in proton-proton (upper row) and proton-neutron
(lower row) collisions. The dashed (dotted) curves depict the contribution
of diagrams with bremsstrahlung from $\gamma\Delta N$ ($\gamma NN$) vertices. In $pn$ channels,
the meson exchange and seagull type diagram are accounted for as well.
The solid lines are the results of calculations of the cross section as coherent sum of all
nucleon and $\Delta $ contributions.}
\label{fig4}
\end{figure}

With these parameters we calculate the invariant mass distribution of di-electrons
produced in $pp$ and $pn$ collisions.
Figure~\ref{fig4} exhibits  the mass distribution  at two values of
the kinetic energy, $1.04$ and $1.25$ GeV. The dotted lines depict the contribution from
pure bremsstrahlung processes from the nucleon lines  including meson exchange
and seagull diagrams for $pn$ reactions (cf.\ Fig.~\ref{fig3}), while the dashed lines
are the contributions of the $\Delta$. 
The solid lines exhibit the coherent sum of nucleon and $\Delta$
contributions. The normalization of the isospin transition matrix has been chosen
as 1. It can be seen that the $\Delta$ contribution dominates in the whole kinematic range , 
except for the region near the kinematical limit, 
where the nucleon current contributions become comparable with $\Delta$ contributions.
It should be noted that, since at considered energies the off-mass shellness 
of $\Delta$ is not large, the contribution  of off-mass shell parameters
$z_i$ is rather small, except in the region near the kinematical limits 
where the behavior of the cross section is slightly modified. However, since the
electromagnetic coupling constants $g_i$ have been adjusted to experimental data together with
off-mass shell parameters \cite{feuster}, we keep  $z_i$ as in Ref.~\cite{feuster}  
for the sake of consistency.
Also note  that the coupling constants $g_{1,2}$ have been fixed at the photon point, 
i.e., at $q^2=0$, while in our case the virtual photon is massive.  
In principle, one may introduce a $q^2$ dependence of the
coupling strengths in form of some phenomenological form factors which  
avoid an unphysical behavior at large virtuality of $\Delta$, i.e., at large $q^2$.

To have an estimate of the role of these parameters for off-mass shell $\Delta$'s,
it is instructive to investigate the invariant mass
distribution in the Dalitz decay of an off-mass shell
particle with all quantum numbers as the $\Delta$
but with different mass $p^2\neq m_\Delta^2$. Such a quantity is often used in
two-step models when calculating di-electrons
from $NN$ collisions \cite{brat08,ernst,ikh_model}, where in a first step a
$\Delta$ like particle is created, then
in the second step it decays into a nucleon and a di-electron.
There are essentially two options for a treatment of such a process:\\
(1) Everything is on-mass shell, i.e.\ in the $\Delta$ Lagrangian, and in the
positive energy  projector operators $(\hat p +m_\Delta)$, Eq.~(\ref{raritaprop}), and
in the spin projection operator $P_{\frac32}(p,m_\Delta)$, Eq.~(\ref{spinproj}),
one  takes the mass parameter as $m_\Delta\Rightarrow m_X$. 
This means that the produced on-mass shell
particle with a mass $m_X\neq m_\Delta$ is nevertheless described by a  
$\Delta$ Lagrangian with the
same coupling constants. This leads merely to a shift in masses
in the expression for the invariant mass distribution and to a corresponding
enlargement of the phase space volume. Evidently,
since in this case $p^2 = m_X^2$ and due to the relation
$\gamma_\alpha (\hat p+m_X) P_{\frac32}^{\alpha\beta}(p,m_X)
=(\hat p+m_X) P_{\frac32}^{\alpha\beta}(p,m_X)\ \gamma_\alpha=0$,
the dependence on $z_i$ drops out in such a treatment (see Eq.~(\ref{theta})).\\
(2) The $\Delta$  is considered off-mass shell, i.e., in the Lagrangian and
projection operators (\ref{spinproj})
and (\ref{raritaprop}) one keeps the mass parameter as
$m_\Delta$, while $p^2\neq m_\Delta^2$ (see \cite{ernst}).
In this case the positive energy projection operator
$(\hat p+m_\Delta)$ does not commute with the spin projection operator  
$P_{\frac32}(p,m_\Delta)$ causing
problems in the treatment of the Rarita-Schwinger  propagator for off-mass shell
particles (see discussion in Ref.~\cite{ourbrem}). 
Obviously, the off-mass shell parameters $z_i$
can now contribute.
The mass distribution of the Dalitz decay with $p^2\neq m_\Delta^2$ 
but with $P_{\frac32}(p,m_\Delta)$
is called the "off-mass shell" Dalitz decay.

By using Eq.~(\ref{intehral}),
the invariant-mass distribution from the decay of a $\Delta$-like particle into a $e^+e^-$
pair with invariant mass $M$ can be presented in the following form:
\be
\frac{d\Gamma^{\Delta_x\to Ne^+e^-}}{dM}=-\frac{\alpha^2_{em}}{12\pi}\frac{|{\bf p}_N|}{m_{X}^2}
\frac1M K_\mu \  K^{+\mu},
\label{dgama}
\ee
where $|{\bf p}_N|$ is the momentum of the outgoing nucleon 
in the $\Delta$ center of mass system,
and the amplitude of a virtual photon production $\Delta_X\to N\gamma^*$ is defined as
\be
 K_\mu \  K^{+\mu}={\rm Tr} \left[
(\hat p_N +m)\ {G^{\mu,}}_\alpha \ (\hat p+m_X) 
P_{\frac32}^{\alpha\beta}(p,m_X)\bar G_{\mu,\beta} \right]
\ee
with
\begin{equation}
G^{\mu,\alpha}=-\left[
\frac{g_1}{2m} \gamma^\nu \Theta^{\mu'\alpha}(z_1)+
\frac{g_2}{4m^2}p_N^\nu \Theta^{\mu'\alpha}(z_2)+\frac{g_3}{4m^2}q^\nu
\Theta^{\mu'\alpha}(z_3) \right]
\left[\phantom{\frac12}\!\! q_{\mu'} g^\mu_\nu - q_\nu g^\mu_{\mu'} \right]\gamma_5 .
\label{vertex}
\end{equation}

\begin{figure}[h]  %
\vskip -9mm
\includegraphics[width=0.475\textwidth]{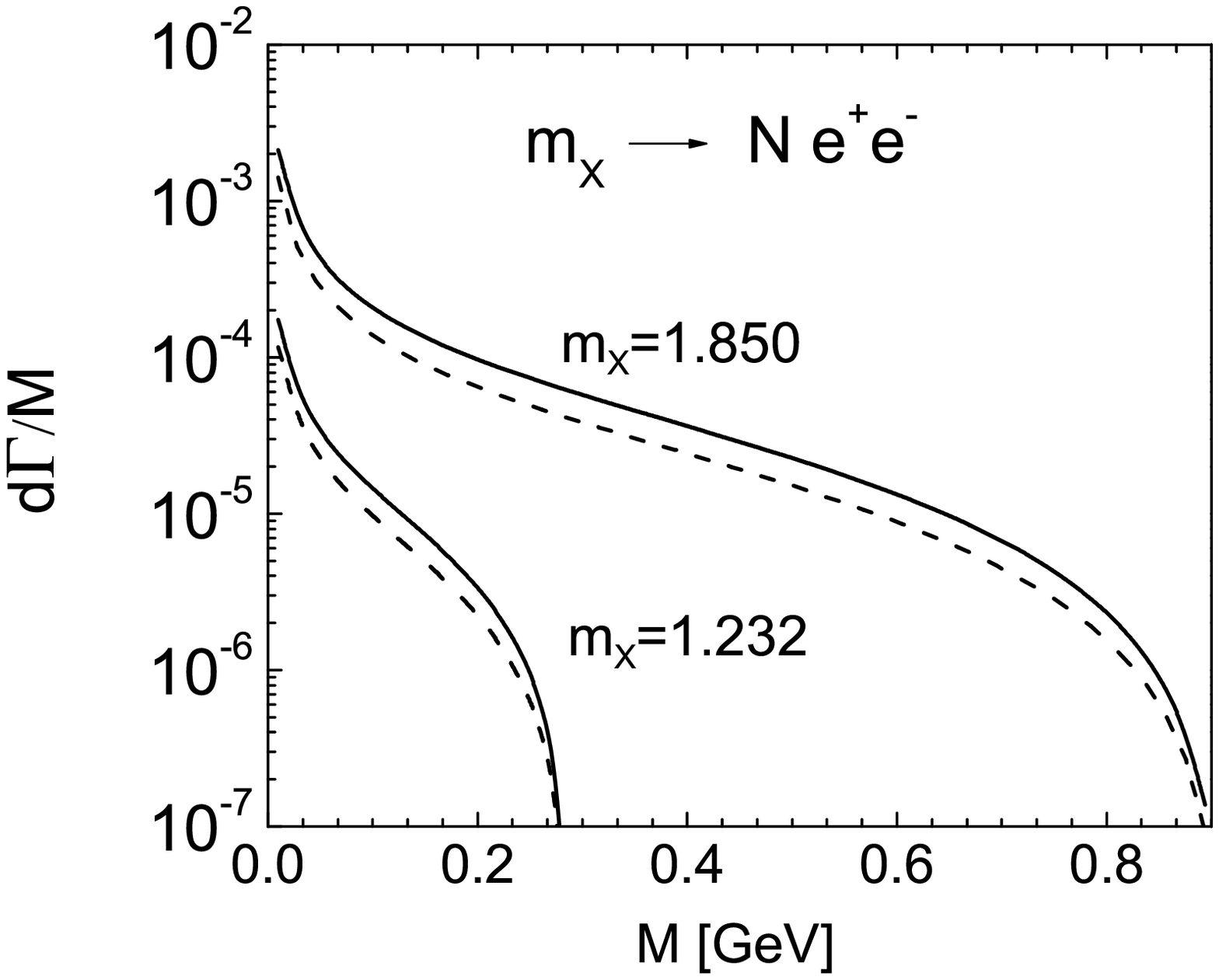} %
\includegraphics[width=0.475\textwidth]{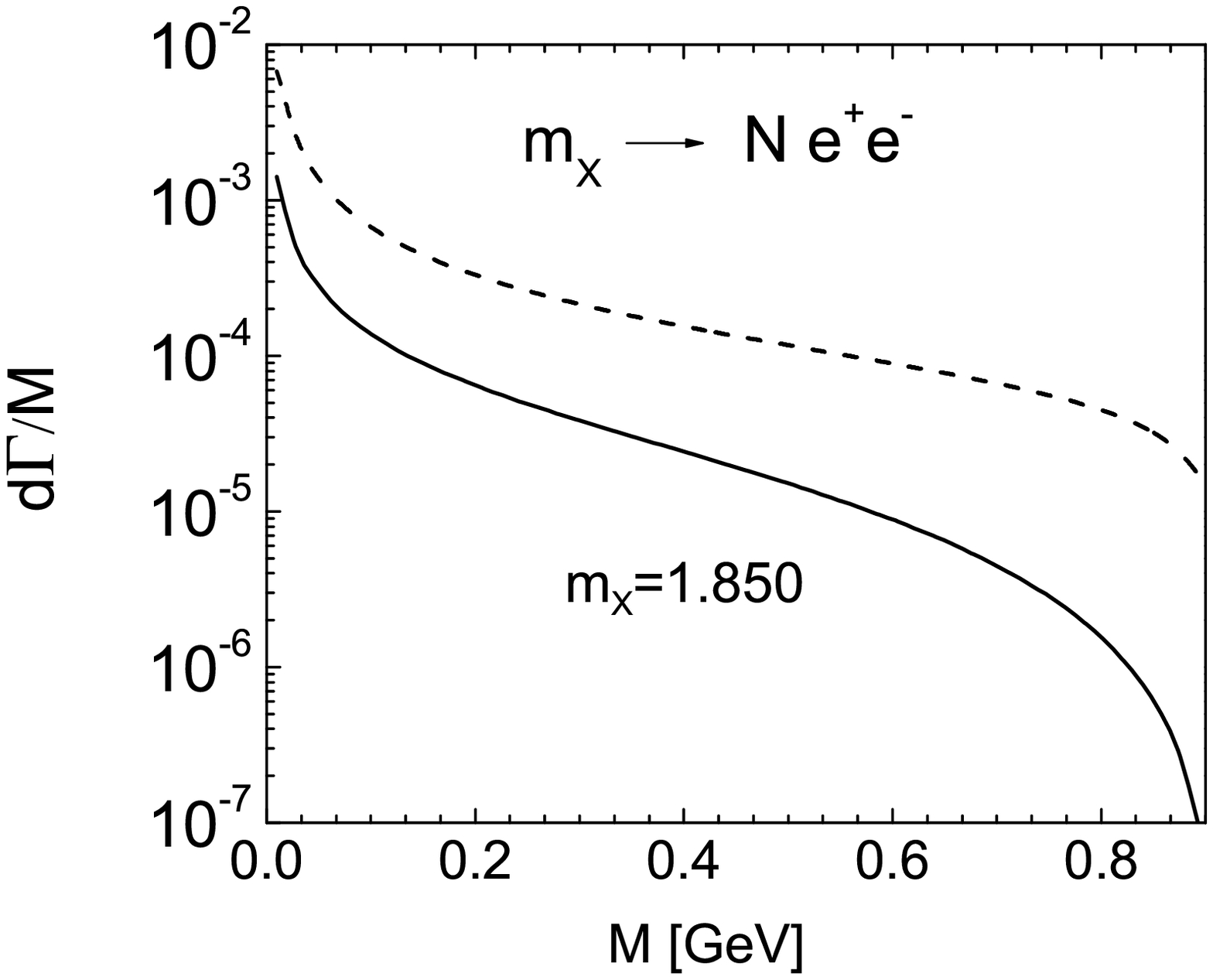} %
\vskip -3mm
\caption{Invariant mass distribution according to Eq.~(\ref{dgama}) 
for the decay of a $\Delta$-like particle
$m_X\rightarrow N e^+e^-$. Left panel: the  mass parameters  
$m_X=1.232$~GeV (lower curves)
and at $m_X=1.850$~GeV (upper curves) are taken on-mass shell, i.e., 
case (1) with $p^2=m_X^2$. The solid and dashed lines correspond to isospin
normalization $\la {\bf T}_3\ra$ to 1 and to $\sqrt{\frac23}$, respectively.
Right panel: The on-mass shell mass distribution (solid line) versus off-mass shell
calculations in case (2), i.e., with the spin projection operator
$P_{\frac32}(p,m_X)$, Eq.~(\ref{spinproj}), 
and the electromagnetic vertices Eq.~(\ref{vertex})
calculated at $p^2\neq m_\Delta^2$ (dashed line). In all calculations the
coupling constants in Eq.~(\ref{vertex}) are
$g_1=5.478$, $g_2=7.611$ and $g_3=7.0$. For the right panel the
off-mass shell parameters have been taken as $z_i=-0.5$.}
\label{distr1}
\end{figure}

In Fig.~\ref{distr1} (left panel), the mass distribution (\ref{dgama})
is presented for the case of on-mass shell $\Delta$-like particles. It can be seen that
due to larger phase space volume, the mass distribution of heavier particles is much larger
than the distribution for real on-mass shell~$\Delta$.

In Fig.~\ref{distr1} (right panel), a comparison
of results of calculations
of the off mass shell distribution are presented for $m_X=1.85$~GeV. The solid line is
for the "on-mass shell" case (1), while the off-mass shell results for case (2)
are presented by dashed lines.
It can be seen that the off-mass shell calculations considerably differ 
from the commonly accepted
results with $\Delta$ on-mass shell. This demonstrates that in two-step models, besides the
traditional approximations, there  are additional uncertainties  in the treatment the
Dalitz decay of the off-mass shell $\Delta$ at large values of invariant mass.
It is clear  that, if in two-step models off-mass shell calculations are employed,  
additional form factors
must be considered to suppress such an increase of the mass distribution 
at large virtuality of the decaying $\Delta$.

\subsection{Spin-$\frac12 $ resonances $P_{11}(1440)$ and $S_{11}(1535)$}\label{spin05}

Contrarily to the $\Delta$ isobars, the isospin-$\frac12$ nucleon resonances 
can couple not only with
isospin-1  mesons, but also $\sigma$, $\eta$ and $\omega$ mesons may contribute.
The corresponding Lagragians are
\begin{eqnarray}
{\cal L}_{NN^*ps}^{(\pm)} &=& \mp \frac{g_{NN^* ps} }{m_{N^*}  \pm m }
\bar\Psi_{N^*}\left\{
\begin{array}{c}
\gamma_5\\ 1\end{array} \right\} \gamma_\mu (\partial^\mu \Phi_{(ps)}) N  + h.c. , 
\label{lps}\\
{\cal L}_{NN^*V}^{(\pm)} &=&  \frac{g_{NN^* V} }{2(m_{N^*}  + m ) }
 \bar\Psi_{N^*}(x)\left\{
\begin{array}{c}
1\\ \gamma_5 \end{array} \right\} \sigma_{\mu\nu} V^{\mu\nu}(x) N  + h.c. , 
\label{lv}\\
{\cal L}_{NN^*\sigma }^{(\pm)} &=& -g_{NN^*\sigma} \bar \Psi_{N^*}  \left\{
\begin{array}{c}
1\\ \gamma_5 \end{array} \right\} N  \it\Phi_{(\sigma)} +h.c.
\label{mnn1}
\end{eqnarray}
with the abbreviations $ps \equiv \pi$ or $\eta$,
$\Phi_{(ps)} \equiv( \boldtau \boldPhi_{(\pi)}) $ or
$\Phi_{(\eta)}(x)$,
$V \equiv V_{(\omega)}$ or
$V(\boldtau\boldrho)$, and
$V^{\alpha\beta}=\partial^\beta V^\alpha - \partial^\alpha V^\beta$.
(The additionally introduced $\eta$ exchange employs for the
$NN\eta$ interaction
${\cal L}_{NN\eta}=
-\frac{f_{NN\eta}}{m_\eta} \bar N\gamma_5 \gamma^\mu (\partial_\mu\Phi_{(\eta)})N$
with $f_{NN\eta} = 1.79$ from Ref.~\cite{bonncd}.)
At a first glance, the consideration of Lagrangians (\ref{lps})-(\ref{mnn1}) involves
into the calculations additional free parameters. However, the effective constants
are not completely free and indeed they can be estimated from independent experimental data.
The coupling constants of the resonances with the pseudo-scalar meson can be found
from the calculations of the experimentally known \cite{PDG}
partial width of the corresponding decay
\begin{eqnarray}
\Gamma_{N^*\to N ps}^\pm=g_{ NN^* ps}^2
\frac{{\cal I}}{8\pi}
\frac{|{\bf p^*}|}{m_{N^*}^2}
\left[ \left( m_{N^*}\mp m \right )^2-m_\pi^2\right],
\label{qqe}
\end{eqnarray}
where ${\cal I}=3$ for pions and ${\cal I}=1$ otherwise.
This yields $g_{ NN^*\pi}^2=6.54$,  $g_{ NN^*\eta}^2=0.5$ for the Roper resonance $N(1440)$
and $g_{ NN^*\pi}^2=1.25$,  $g_{ NN^*\eta}^2=2.02$ for the $N(1535)$ resonance.

\begin{figure}[h]  %
\vskip -3mm
\includegraphics[width=0.5\textwidth]{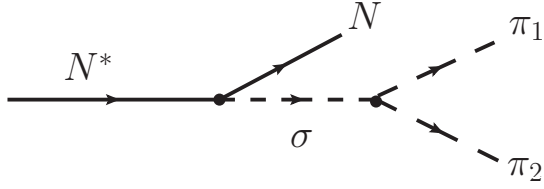} %
\caption{Diagram for the decay of a resonance $N^*$ in to two pions 
in the state $S=0$, $I=0$ with an intermediate $\sigma$ meson.}
\label{fig11}
\end{figure}

Estimates of coupling constants with $\sigma$ mesons, 
for which direct experimental data are not available, are more involved.
One of the method is to calculate the  decay of the considered resonances 
into two pions in a specific
final state, $N (2\pi)^{I=0}_{S=0}$. For low-lying resonances 
such a decay can be treated as a process with intermediate
excitation of the $\sigma$ meson with its subsequent decay into two pions \cite{soyer2pi}, 
as depicted in Fig.~\ref{fig11}.

A direct calculation of the diagram Fig.~\ref{fig11} results in
\be\!\!\!\!\!\!\!\!\!
\Gamma_{N^*\to N(2\pi)^{I=0}_{S=0}}^\pm=\frac{g_{NN^*\sigma}^2}{4\pi}
\int\limits_{2m_\pi}^{m_{N^*}-m} d\xi
\frac{|{\bf p^*}|m_\sigma^2\left[ (m_{N^*}\pm m)^2-\xi^2\right] }
{|\xi^2-(m_\sigma-\frac{i}{2} \Gamma)^2|^2}
 \frac{\Gamma_{\sigma\to 2\pi}}{\pi\ m_{N^*}^2}
\sqrt{ \frac{\xi^2-4m_\pi^2}{m_\sigma^2-4 m_\pi^2}},
\ee
where the dependence of the width $\sigma\to 2\pi$ on the mass $\xi$ of the intermediate meson
is computed from the Lagrangian
\be
{\cal L}_{\sigma\pi\pi}=g_{\sigma\pi\pi}\frac{m_\sigma}{2}
\left ( \boldPhi_{(\pi)} \boldPhi_{(\pi)} \right)\Phi_{(\sigma)}
\label{sigmapipi}
\ee
with the coupling constant $g_{\sigma\pi\pi}$  found from the total decay width of the
$\sigma$ meson into two pions.
In the present calculation we adopt $m_\sigma=500$~MeV and 
$\Gamma_{\sigma\to 2\pi}=250$~MeV being
consistent with the recent analysis \cite{leutw}. With these parameters
we obtain $g_{NN^*\sigma}=2.1$ for the Roper resonance and   
$g_{NN^*\sigma}=3.8$ for
$N(1535)$ (see also \cite{nakayamac69}). 
The propagator of the off-mass shell resonance, $p_X^2\neq m_{N^*}^2$,
is augmented by a form factor of the form
\be
F(p_X)=\frac{\Lambda^4}{\Lambda^4+(p_X^2-m_{N^*}^2)^2 }
\ee
with $\Lambda=1.2$~GeV in both cases, for Roper and $N(1535)$ resonances.
The finite widths of the intermediate resonances are taken into account as usually:
adding to the mass parameter in the
propagator an imaginary part as $m_{N^*}\to m_{N^*} -i\Gamma^{tot}/2$.
In our case, the resonances are off mass shell and their total widths depend on
the invariant masses of the resonance. Such a dependence can be parameterized as \cite{feuster}
\be
\Gamma^{tot}(m_X)=\sum_i \Gamma_i(m_X) F_X(m_X),
\label{gamaote}
\ee
where the sum runs over all  possible partial decay channels, and
the cut-off form factors $F_X(m_X)$ suppress an unphysical increase of the width with
increasing $m_X$.
\be
 F_X(m_X)=\frac{2}{1+\left( \frac{P^*(m_X)}{P^*(m_{N^*})}\right)^\alpha},
\ee
where $P^*$ is the nucleon momentum in the resonance center of  mass system,
 $\alpha=3$  and $\alpha=2$ for the  Roper and $N(1535)$ resonances, respectively.
For the Roper resonance there are two main decay channels, $N(1440) \to N \pi$
(branching ratio $\sim 60- 70 \% $)
and $N(1440) \to \Delta \pi$ (branching ratio $\sim 20- 30 \%$) \cite{PDG}, 
while $N(1535)$ decays mainly either into a nucleon
and a pion (branching ratio $\sim 55\%$) or into a nucleon and 
$\eta$ (branching ratio $\sim 45\%$).
The energy dependence of the partial widths $\Gamma_i(m_X)$ 
for two-body decay have been calculated using the same
Lagrangians (\ref{lps})-(\ref{mnn1}). 
For the decay of the Roper resonance  $N^*\to N \pi\pi$ via $\Delta$ we
employ an   effective Lagrangian of the form
\be
{\cal L}_{ N^*\Delta\pi} =
\frac{f_{N^*\Delta\pi}}{m_\pi} \bar \Psi^\alpha {\bf T} \partial_\alpha
\boldPhi_{(\pi)} N^* + h.c..
\label{ldelta}
\ee
Note that by calculating the energy dependence of the widths 
from the Lagrangian (\ref{ldelta}) a knowledge
of the coupling constant $f_{N^*\Delta\pi}$ is not necessary.
Also note that, in spite of the mass of the Roper resonance which is only slightly above
the kinematical limit,
the probability to decay into $\Delta$ and $\pi$ is relatively large. This is due
to the large total width of the $\Delta$ resonance which  correspondingly spreads the mass
around the pole position, enlarging therefore the phase space of the decay channel. 
In calculating the partial width $N^*\to \pi \Delta$ we adopt a Gaussian  distribution of
the mass of the $\Delta$ resonance
$f(\tilde m_{\Delta}) =  
(\sigma_\Delta\sqrt{2\pi})^{-1} \exp[-(\tilde m_{\Delta}-m_\Delta)^2/(2\sigma_\Delta^2)]$
with $\sigma_\Delta = \Gamma_\Delta$.
The remaining coupling constants for the vector mesons have been taken from 
Ref.~\cite{nakayamac69} (see Tab.~\ref{tablitza}) .

\begin{table}[h]
\caption{Coupling constants $g_{NN^*M}$ and cut-off parameters $\Lambda$ for the effective
Lagrangians (\ref{lps})-(\ref{mnn1}), computed either directly from the
decay widths or taken from Ref.~\cite{nakayamac69}.}

\begin{tabular}{c c c c ccc }
\hline\hline 
Meson    &\phantom{p} & $P_{11}(1440)$  &\phantom{q}& $S_{11}(1535)$& &$\Lambda$ [GeV]\\\hline
$\pi $   &            & 6.54            &           & 1.25          & &   1.2\\
$\eta$   &            & 0.5             &           & 2.02          & &   1.2\\
$\sigma$ &            &  2.1            &           & 3.8           & &   1.2\\
$\rho$   &            & -0.57           &           & - 0.65        & &   1.2\\
$\omega$ &            &- 0.37           &           & -0.72         & &   1.2\\
\hline\hline
\end{tabular}
\label{tablitza}
\end{table}

The effective Lagrangian for the
electromagnetic decay of the resonance into a photon and a nucleon has been taken as
\be
{\cal L}_{NN^*\gamma}^\pm=\frac{e\kappa}{2m_R}\bar \Psi_{N^*}
\left\{
\begin{array}{c}
1 \\
\gamma_5
\end{array}\right \}
\sigma^{\mu\nu}N F_{\mu\nu}+h.c.,
\ee
where, in contrast to the $\Delta$ case, the coupling
$\kappa$ is different for proton and neutron vertices and 
can be found from the helicity amplitudes
\be
A_{1/2}=-{e\kappa}\ \sqrt{\frac{|{\bf p^*}|}{m m_{N^*}}},
\ee
where for the Roper resonance one has  $A_{1/2}^p=-0.065 \ \rm GeV^{-1/2}$ and
$A_{1/2}^n=0.04 \ \rm  GeV^{-1/2}$, while for the
$N(1535)$ we employ $  A_{1/2}^p=0.09\  \rm GeV^{-1/2}$ 
and $A_{1/2}^n=-0.046\ \rm GeV^{-1/2}$, correspondingly.

\subsection{Spin-$\frac32 $ nucleon resonances}\label{spin15}

The next considered resonance is $D_{13}(1520)$ with negative parity and spin $\frac32$.
The effective Lagrangian for such resonances is chosen in the same form as for the 
$\Delta$ with the exception that, since the isospin is $\frac12$, 
besides the isovector mesons $\pi$ and $\rho$, the isoscalar $\eta$, $\omega$ 
and $\sigma$ also can contribute. The effective Lagrangians are as follows
\begin{eqnarray}
{\cal L}_{NN^*ps}^{(\pm)} &=&  \frac{g_{NN^* ps} }{m_{ps}}
 \bar\Psi_{N^*}^\alpha(x)\left\{
\begin{array}{c}
1\\ \gamma_5\end{array} \right\}  (\partial_\alpha \Phi_{(ps)})  N + h.c. ,\\
{\cal L}_{NN^*V}^{(\pm)}& = & \mp i\frac{g_{NN^* V} }{m_V }
 \bar\Psi_{N^*}^\alpha
\left\{\begin{array}{c}\gamma_5\\ 1\end{array} \right\}
\gamma^{\lambda} V_{\alpha\lambda} N  + h.c. ,\\
{\cal L}_{NN^*\sigma}^{(\pm)} &=&  i\frac{g_{NN^* \sigma} }{m_{\sigma}}
 \bar\Psi_{N^*}^\alpha \left\{
\begin{array}{c}
\gamma_5\\ 1\end{array} \right\}  (\partial_\alpha \Phi_{(\sigma)})  N + h.c..
\end{eqnarray}
Note that in choosing the relative phase for the $\sigma$ meson 
an imaginary unit $i$ must be explicitly
displayed. In principle, to synchronize the relative phases of different Lagrangians 
one may compute the corresponding amplitude in a fully coplanar kinematics. 
Then such an amplitude, in tree level calculations,
must be either purely real or purely imaginary.
The coupling constants and the cut-off parameter $\Lambda=0.8$~GeV 
have been taken from Ref.~\cite{mosel_calc}.
The propagator is chosen in the form (\ref{raritaprop}) 
with the resonance mass augmented by the total
decay width. The total width of the resonance $N(1520)$ is calculated by Eq.~(\ref{gamaote})
for which three decay channels
have been taken into account, $N(1520) \to N \pi$ (branching ratio $\sim$ 50\%),
$N(1520) \to N \rho$ (branching ratio $\sim$ 25\%) and
$N(1520) \to N \Delta$ (branching ratio $\sim$ 25\%). For the decay in to $N\rho$
and $N\Delta$ the partial widths
have been calculated again by adopting Gaussian distributions of
the mass  of $\rho$  and $\Delta$ around their pole values.

The electromagnetic part of the Lagrangian has the same form as for the $\Delta$, 
see Eq.~(\ref{deltaLag}), except
for the isospin transition matrix ${\bf T}$, i.e.
\begin{eqnarray}
{\cal L}_{N^* N \gamma}^\pm&=& -i\frac{eg_1}{2m}
\bar\Psi^\alpha \Theta_{\alpha\mu}(z_1)\gamma_\nu
\left\{\begin{array}{c}
\gamma_5 \\ 1
\end{array}\right\}
N {\cal F}^{\nu\mu}-  
\frac{eg_2}{4m^2}\bar\Psi^\alpha \Theta_{\alpha\mu}(z_2)
\left\{\begin{array}{c}
\gamma_5 \\ 1
\end{array}\right\}
\left( \partial_\nu N \right) {\cal F}^{\nu\mu} \nonumber\\
&& -\frac{eg_3}{4m^2}\bar\Psi^\alpha \Theta_{\alpha\mu}(z_3)
\left\{\begin{array}{c}
\gamma_5 \\ 1
\end{array}\right\}
N  \partial_\nu {\cal F}^{\nu\mu} +h.c.. \label{Lag1520}
\end{eqnarray}
\noindent
As in case of $\Delta$, the two coupling constants $g_{1,2}$ can be obtained from the helicity
amplitudes, separately for proton and neutron. 
The remaining constants $g_3$, $z_1,z_2$ and $z_3$, being completely
free parameters,
are to be found by  fitting experimental data. There exist several sets in the literature
equally well describing the corresponding data \cite{feuster}. 
In our calculations we have chosen
$g_1=3.004, g_2=3.047,\ g_3=0$ for the proton and
$g_1=-0.068, g_2=1.265,\ g_3=0$ for the neutron. The off-mass shell parameters
$\ z_1=1.39$ and $z_2=0.267$ have been also taken from \cite{feuster}
(since $g_3 = 0$, the off mass shell parameter $z_3$ is irrelevant here).

\begin{figure}[h]  %
\vskip -9mm
\hspace*{-9mm}\includegraphics[width=1.1\textwidth]{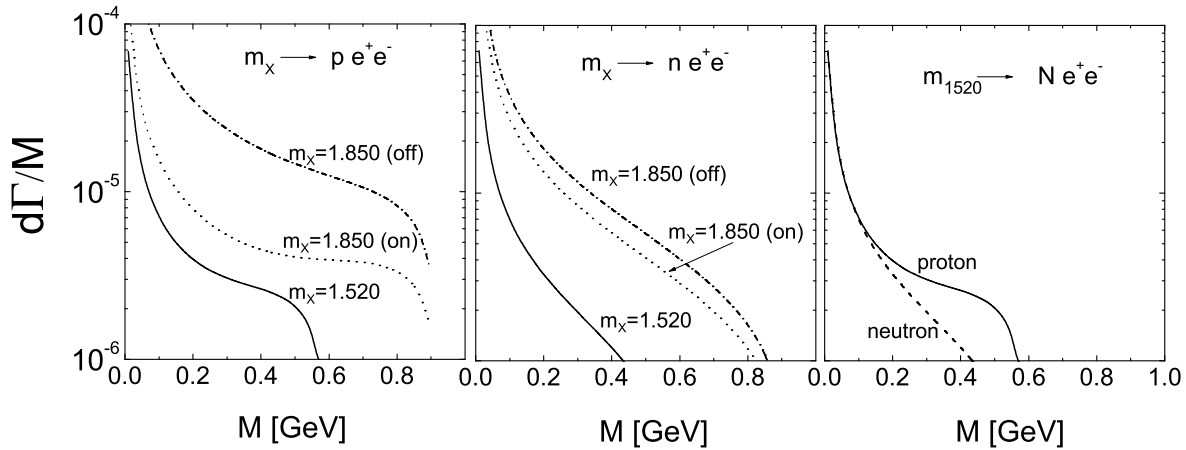} %
\vskip -9mm
\caption{Invariant mass distribution according to Eq.~(\ref{dgama}) for
the decay of a $N(1520)$-like particle
$m_X\rightarrow N e^+e^-$ into a proton (left panel) and a neutron (middle  panel)
at two values of the  mass parameter $m_X=1.520$~GeV and $m_X=1.850$~GeV.
For the case of $m_X=1.850$~GeV two different definitions of
$d \Gamma^{\Delta\to Ne^+e^-} / d M$
have been used corresponding to cases (1) and (2) in subsection \ref{isobar}
(see text there).
The solid and dotted lines correspond to the case (1), where the resonance is on
mass shell; the dot-dashed lines are results of calculations for case (2)
with the resonance off mass shell, i.e., in the spin projection operator
$P_{\frac32}(p,m_X)$, Eq. (\ref{spinproj}), and the electromagnetic vertices (\ref{vertex})
the mass parameter is $m_{X}=1.520\ \rm GeV$, while $p^2=1.850\ \rm GeV$. In the right panel,
the  comparison of the  mass distribution for proton and neutron vertices reflects the
relative contribution of the $N(1520)$ resonance in $pp$ and
$pn$ di-electron production (cf.\ solid curves in the left and middle panels).}
\label{Brn1520}
\end{figure}

As in case of $\Delta$ isobar we calculate the mass distribution 
(\ref{dgama}) of the Dalitz decay of
a particle with quantum numbers of $N(1520)$ at different masses. This distribution is also
frequently used in calculations by two-step models \cite{brat08}.
Figure \ref{Brn1520} illustrates the behavior of the invariant mass distribution
in the Dalitz decay of a $D_{13}$-like resonance at different masses 
(left and middle panel) for the
calculations of on-mass and off-mass shell decays. 
It can be seen that, as in the case of $\Delta$ isobar
(cf.\ Fig.~\ref{distr1}),  the heavier masses result in  broader distributions. The virtuality
of the decaying  resonance leads to an increased  broadening  distribution. Also,
since the neutron couplings $g_{1,2}$ are much smaller than the proton ones, 
the effect of virtuality
is less pronounced  in the neutron case. In the right panel of
Fig.~\ref{Brn1520} we present a direct comparison
of the contribution of the $N(1520)$ resonance for proton and neutron decays. 
At larger di-electron
invariant masses the relative contribution of the $N(1520)$  
in $pn$ collisions is smaller than
in $pp$ processes.

\begin{figure}[h]  %
\vskip -12mm
\includegraphics[width=1.\textwidth]{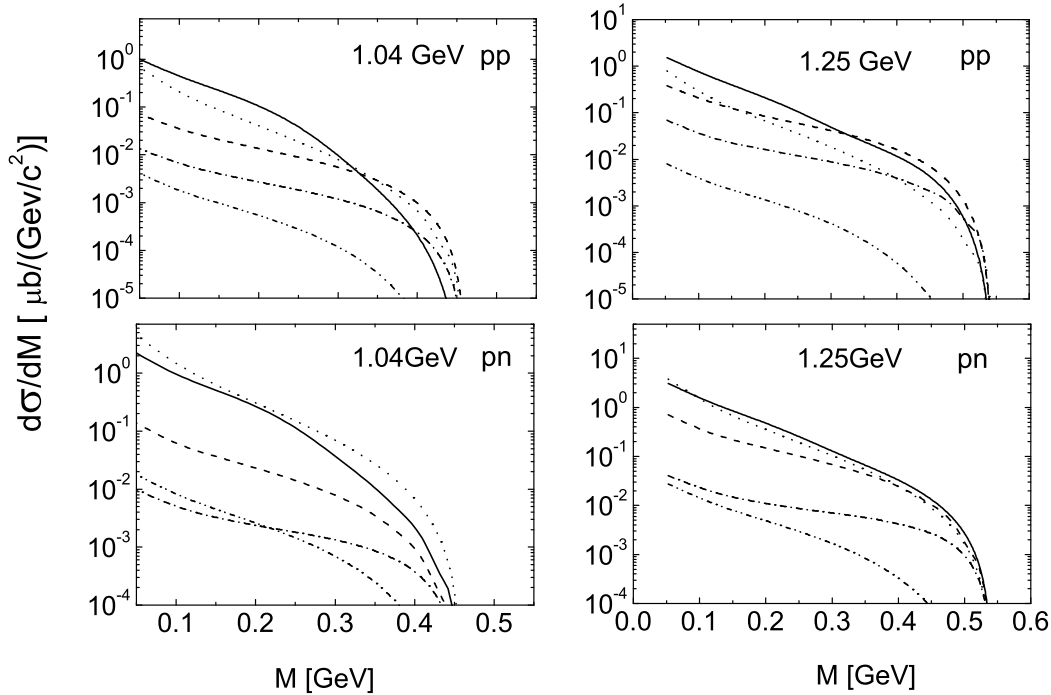} %
\vskip -9mm
\caption{Invariant-mass distribution of $e^+e^-$ in $pp$ (top) and
$pn$ (bottom) collisions. 
The dotted curves depict the contribution of diagrams with bremsstrahlung
from $\gamma NN$ vertices, cf. Fig.~\ref{fig3}.
Dash-dot-dot lines: contribution of Roper-resonance, 
dash-dotted lines: $N(1535)$,
dashed lines: $N(1520)$. 
Solid curves are for all resonances including the contribution from $\Delta$.}
\label{resonances}
\end{figure}

With the above parameters we calculate  the contribution
of the mentioned baryon resonances  in the invariant mass distribution
of di-electrons from $pp$ and $pn$ collisions.
In Fig.~\ref{resonances} we present the comparison of the nucleon
bremsstrahlung contribution with  individual contributions
from each of the considered resonances
in $pp$ (upper row) and $pn$ (lower row) collisions at two kinetic energies.

The role of the Roper resonance (dash-dot-dot lines)  is negligibly small in all cases.
Also a small contribution comes from $N(1535)$ (dash-dotted lines) which becomes of the
same order as the nucleon bremsstrahlung
only at the kinematical limit. A more significant
contribution stems from the $N(1520)$ resonance, which becomes
competitive with nucleon bremsstrahlung already at di-electron invariant mass 
$M\ge 0.45$ GeV.
The solid line is the coherent sum of all the resonances, including $\Delta$ isobars.
In these calculations and in what follows, 
the isospin transition matrix has been normalized to $\sqrt{2/3}$.
The contribution of $N(1535)$ becomes more pronounced at di-electron invariant masses 
corresponding to the pole position of the resonance mass. 
Note that in our calculations the initial state interaction
has been taken into account by imposing a energy dependence of the effective parameters 
as suggested in Ref.~\cite{mosel_calc}. 
In principle, one could account for the effects of initial state
interaction explicitly, as proposed in \cite{nakayamaISI}. An analysis performed in  
\cite{shyam08} shows that at the considered energies the two methods of accounting 
for the initial state effects provide similar results. 
The effects of final state interaction (FSI)
in the considered reactions have been investigated in
Ref.~\cite{ourbrem}. It has been found that FSI corrections depend 
on the relative momentum of the outgoing nucleons,
becoming significant at low momenta, i.e. 
at the kinematical limit of the di-electron invariant mass.
At low and intermediate values of the di-electron mass FSI effects are small.

\begin{figure}[h]  %
\vskip -12mm
\hspace*{-.5cm}\includegraphics[width=1.1\textwidth]{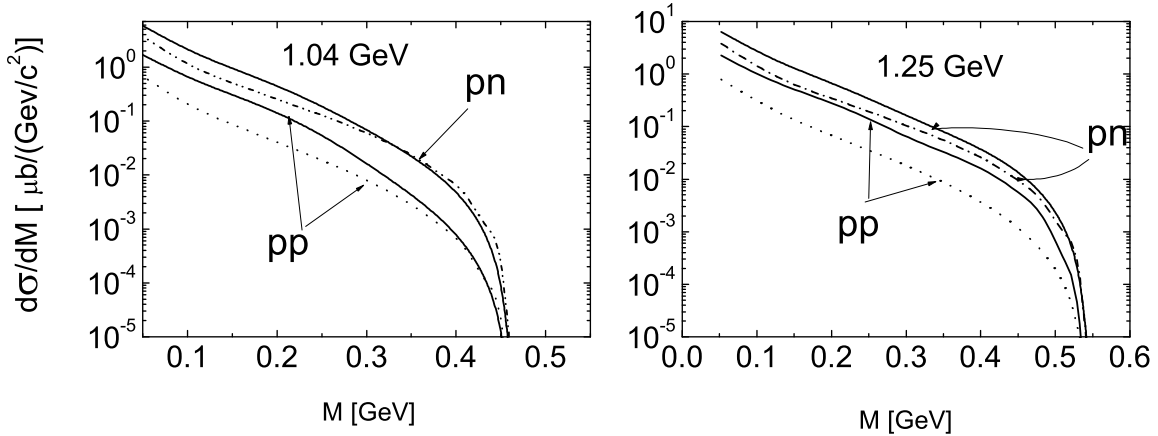} %
\vskip -12mm
\caption{Invariant mass distribution of $e^+e^-$ pairs in $pp$  and
$pn$ collisions at two kinetic energies (left: 1.04 GeV, right: 1.25 GeV). 
The dotted (dash-dotted) lines depict
contributions of bremsstrahlung
diagrams  without resonances in $pp$ ($pn$) collisions. 
The solid lines are the results of the coherent sum
of all the diagrams, including bremsstrahlung and contributions from
$P_{33}(1232)$, $P_{11}(1440)$, $D_{13}(1520)$ and $S_{11}(1535)$ resonances.}
\label{total}
\end{figure}

Eventually, the total cross section with accounting for all resonances and 
nucleon bremsstrahlung is presented in Fig.~\ref{total} by solid lines. 
The dotted and dash-dotted curves exhibit the contribution
from the nucleon bremsstrahlung solely in $pp$ and $pn$ collisions respectively. 
One concludes from Fig.~\ref{total}
that the contribution of baryon resonances dominates the cross section at the
considered energies.

\begin{figure}[h]  %
\vskip -12mm
\hspace*{-0.5cm}\includegraphics[width=0.7\textwidth]{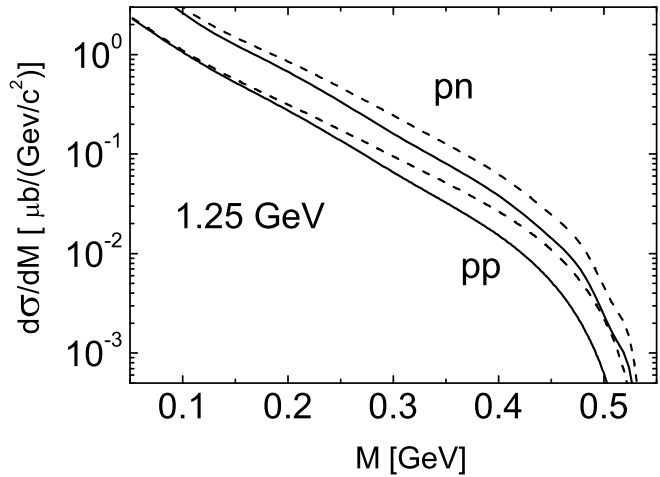} %
\vskip -9mm
\caption{Invariant mass distribution of $e^+e^-$ in $pp$ and
$pn$ collisions as a coherent sum of the considered Feynman diagrams (solid lines)
vs. an incoherent summation (dashed lines) of separate contributions from bremsstrahlung,
$P_{33}(1232)$, $P_{11}(1440)$,  $D_{13}(1520)$ and $S_{11}(1525)$, respectively.}
\label{interference}
\end{figure}

It is worth emphasizing that in the present calculations 
the quantum mechanics interference effects
play an important role in the total cross section, 
essentially reducing the cross section in comparison to
a incoherent sum of different contributions. 
This is illustrated in Fig.~\ref{interference}, where
results of a coherent summation of  Feynman diagrams are presented and compared with
the incoherent sum of separate contributions of bremsstrahlung and
$P_{33}(1232)$, $P_{11}(1440)$, $D_{13}(1520)$ and $S_{11}(1535)$
resonances. It is seen that in both cases, $pn$ and $pp$ collisions,
the interference effects become significant at higher values of the di-electron
invariant mass and reduce the cross section by a factor of about $2 - 2.5$.

Experimentally, information on the di-electron production from $pn$ collisions may be extracted
from the tagged neutrons in $Dp\to p_{sp}np \ e^+e^-$ reactions 
by exploiting the so-called spectator
mechanism. As discussed in Ref.~\cite{ourbrem}, 
if the spectator  proton  $p_{sp}$ is detected in the very
forward direction with about half of the momentum of the incident deuteron, 
then  with a high probability
the reaction occurred at the neutron, and the proton from the deuteron remains as a spectator.
In  such a case, one may extract the $pn$ sub-reaction at the same energy 
as the detected proton.
To reduce the experimental errors one may measure the ratio $\sigma_{pn}/\sigma_{pp}$ in such
experiments. However, even such a ratio may remain rather sensitive 
to the extraction procedure, namely to the
accuracy of determining the effective momentum of the tagged active neutron.

\begin{figure}[h]  %
\vskip -9mm
\hspace*{-0.5cm}\includegraphics[width=0.6\textwidth]{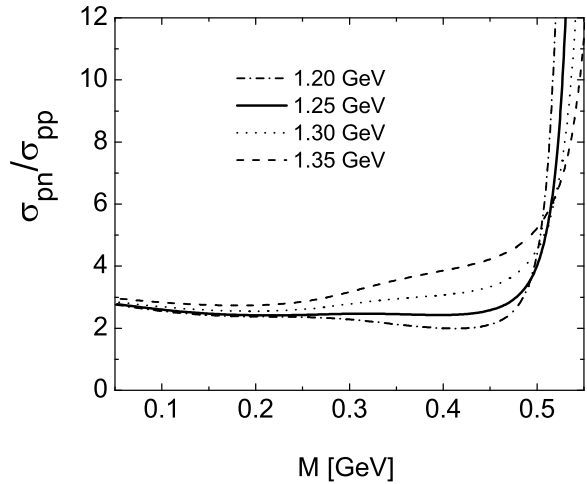} %
\vskip -9mm
\caption{Ratio of the  $e^+e^-$ invariant mass distribution in $pn$
and $pp$ collisions at kinetic beam energy of 1.25 GeV for the $pp$  reaction
and four different energies for the $pn$ reaction:
1.20 GeV (dash-dotted line), 1.25 GeV (solid line), 1.30 GeV (dotted line)
and 1.35 GeV (dashed line).}
\label{ratio}
\end{figure}

In Fig.~\ref{ratio} we present the ratio $\sigma_{pn}/\sigma_{pp}$ 
calculated at few different kinetic energies in the $pn$ collisions 
while keeping fixed the kinetic beam energy of 1.25 GeV for the $pp$ reaction.
Such a ratio emulates roughly the possible Fermi motion
effects in the $Dp\to p_{sp}np \ e^+e^-$ subreaction. It can be seen that, for 
di-electron invariant masses $M > 300$ MeV, 
the presented ratio is quite sensitive to 
the effective momenta of the neutron.

\section{Summary} \label{summary}

In summary
we have analyzed various aspects of the di-electron production from the
bremsstrahlung mechanism and resonance excitations at intermediate energies for
the exclusive reactions $NN\to NN\: e^+e^-$,
i.e., for $pp \to pp e^+ e^-$ and  $np \to np e^+ e^-$.
To calculate the corresponding cross sections we employ an effective meson-nucleon theory 
with parameters adjusted to elastic $NN$ and inelastic $NN\to NN\pi$
reaction data with low-mass baryon resonances included.

The performed evaluations of bremsstrahlung diagrams can be considered
as an estimate of the background contribution, a detailed knowledge of which is a necessary
prerequisite for understanding di-electron production in heavy-ion collisions.
Our approach is based on covariant evaluations of the corresponding
tree level Feynman diagrams with implementing phenomenological form factors,
with particular attention paid on preserving the gauge invariance. It is stressed that,
regardless of the choice of the pion-nucleon-nucleon coupling, the consideration of
seagull type diagrams is inevitable if meson field derivatives enter the interaction 
Lagrangians, say for $\rho$ mesons. 
The covariance of the approach is ensured by direct calculations of Feynman diagrams.

In accordance with previous results \cite{mosel_calc,ourbrem} our calculations demonstrate
that in the region of invariant masses sufficiently far from the vector meson production
threshold the main contribution to the cross section, in both reactions $pp$ and $pn$,
comes from virtual excitations of  nucleon resonances. The contribution from the Roper
resonance is negligibly small in all the considered reactions. 
The role of the $S_{11}(1535)$ resonance is also marginal in the whole kinematical region 
except for values of the invariant mass at the kinematical limit. 
The main contribution to the cross sections comes from spin-$\frac32$ resonances, 
$\Delta$ and $N(1520)$, 
the role of the latter increasing  with increasing  initial $NN$ energy.
Due to isospin effects and meson exchange diagrams the cross
section for the reaction $pn\to pn\: e^+e^-$  is larger than the cross section
for  $pp\to pp\ e^+e^-$ by a factor $ 1.5-3$. Note that because of
(i) contributions
of the isoscalar $\sigma$ and $\omega$ exchange mesons,
(ii) differences
in the electromagnetic coupling in $\gamma p$ and $\gamma n$ systems,
(iii) interference effects, and
(iv) contribution of resonances,
the isospin enhancement is not $\sim 9$, as one could naively expect from isospin
symmetry considerations. In both reactions, $pn\to pn\, e^+e^-$ and $pp\to pp\ e^+e^-$,
the bremsstrahlung cross sections exhibit a smooth behavior as a function of the
di-electron mass. Hence, the bremsstrahlung cross section
can indeed be considered as background contribution.

The previous ''DLS puzzle'', experimentally resolved in \cite{HADES1},  
seems to be shifted now to a "theory puzzle": 
the preliminary data for the invariant mass spectrum in the reaction
$np\to np\  e^+e^-$, extracted from the tagged subreaction in $Dp\to p_{sp}np \ e^+e^-$,
point to a shoulder at intermediate values of the di-electron invariant mass \cite{panic}.
Such a structure is hardly described within the present approach. 
(In contrast, the use of the phenomenological
one-boson exchange model for the exclusive reactions $NN\to NNM$ 
with $M=\omega, \phi, \eta$ and $\eta'$
\cite{ourOmega,ourPhi,titov,ourEtas}   
successfully describes data and has some prediction power \cite{cracow}).
Thus, understanding the elementary channels remains challenging.
Finally, it should be emphasized that we  consider here the exclusive reaction 
$NN\to NN\, e^+e^-$.
The inclusive channels $NN \to NN \ X \ e^+e^-$ may be significantly different
(cf.\ \cite{gale}).

\section{Acknowledgements}

Discussions with E.L.\ Bratkovskaya, U.\ Mosel, K.\ Nakayama,
B.\ Ramstein, A.I.\ Titov, W.\ Weise and G.\ Wolf are gratefully acknowledged.
L.P.K. would like
to thank for the warm hospitality in the Research Center Dresden-Rossendorf.
This work has been supported by
BMBF grant 06DR136, GSI-FE and the Heisenberg-Landau program.

\end{document}